\documentclass[mathleft
]{an}
\usepackage{graphicx}
\usepackage{times}
\usepackage{natbib}
\begin{document}

\Pagespan{1}{}% Document's page range.
% If second parameter is left empty, the last page is computed automatically.
\Yearpublication{2007}%
\Yearsubmission{2007}%
%\Month{xx}%   
\Volume{328}%  
\Issue{10}% 
\Pagespan{1087}{1091}
\DOI{10.1002/asna.200710836}% 

\title{Are solar cycles predictable?}

\author{M. Sch{\"u}ssler}

\titlerunning{Are solar cycles predictable?}
\authorrunning{M. Sch{\"u}ssler}
\institute{
Max Planck Institute for Solar System Research,
37191 Katlenburg-Lindau, Germany}

\received{2007 Jul 19}
\accepted{2007 Oct 10}
\publonline{2007 Dec 15}


\abstract{Various methods (or recipes) have been proposed to predict
future solar activity levels - with mixed success. Among these, some
precursor methods based upon quantities determined around or a few years
before solar minimum have provided rather high correlations with the
strength of the following cycles. Recently, data assimilation with an
advection-dominated (flux-transport) dynamo model has been proposed as a
predictive tool, yielding remarkably high correlation coefficients.
After discussing the potential implications of these results and the
criticism that has been raised, we study the possible physical origin(s)
of the predictive skill provided by precursor and other methods. It is
found that the combination of the overlap of solar cycles and their
amplitude-dependent rise time (Waldmeier's rule) introduces correlations
in the sunspot number (or area) record, which account for the
predictive skill of many precursor methods. This explanation requires no
direct physical relation between the precursor quantity and the dynamo
mechanism (in the sense of the Babcock-Leighton scheme or otherwise).}

\maketitle

\section{Introduction}

Taken at face value, the question posed in the title has to be answered
in the affirmative: one cannot deny that there is on the market a whole
lot of predictions of future solar activity levels. A quick (and
unsystematic) search in the Smithonian/NASA Astrophysical Data System
(ADS) reveals that there are 50\% more hits for the combination of
"prediction" and "solar cycle" in title or abstract than for "solar
dynamo". In fact, there are alone 281 such hits for publications since
2004 (status: July 16, 2007).  Interestingly, the intersection of both
sets, i.e., papers dealing both with solar-cycle prediction {\em and}
with the solar dynamo comprises less than 5\% of the papers on
prediction. This is not very surprising because, until recently, solar
dynamo models have not been considered to have reached a state of
maturity to be used for predictive purposes.

Most prediction methods in the literature can be categorized into one of
two classes \citep[cf.][]{Wilson:1994}:
\begin{enumerate}
\item {\em Extrapolation methods,} based on statistics 
  or pattern re\-cognition: most relevant information about the system is
  assumed to be contained in the available data (e.g., the
  sunspot number record), so that the future can be extrapolated from
  the past. The simplest example is harmonic analysis
  \citep[e.g.,][]{Echer:etal:2004}, but also concepts of nonlinear
  dynamics are used \citep[e.g.,][]{Sello:2001}.
\item {\em Precursor methods,} assuming that certain physical quantities
  measured during the descending or minimum phase of an activity cycle
  contains information about the strength of the next cycle 
  \citep[e.g.,][]{Lantos:Richard:1998,Hathaway:etal:1999,Schatten:2003}. 
\end{enumerate}
The overall success of the various methods in predicting the future is
rather limited (e.g., see Figure~14.2 in Wilson 1994 and Figure~6 of
Lantos \& Richard 1998). However, if the historical record of solar
activity is considered, some precursors show remarkable levels of
correlation with the strength of the following cycle. For instance,
\citet{Ohl:1966} took the minimum level of geomagnetic variations (as
measured, for instance, by the {\em aa} index) as a precursor for the
strength of the next cycle. This method does not have any adjustable
parameters and yet provides a correlation coefficient of $r=0.91$ for
solar cycles 12-22 \citep{Hathaway:etal:1999}. The method of
\citet{Thompson:1993}, which is also based upon geomagnetic variations,
even yielded $r=0.97$ for the same cycles, but utterly failed in actually
predicting cycle 23: the predicted value for the sunspot number of
$R=160$ turned out to be more than 30\% too high! This result reminds us
that a high correlation coefficient for post-dicting the past does not
necessarily imply a high skill of the method for predicting the future.

\section{Solar cycle predition and dynamo models}

The paper by \citet{Schatten:etal:1978} is (to my knowledge) the first
paper in which the words "dynamo theory" and "sunspot number prediction"
appear together in the title. The authors argue that, in the framework
of the Babcock-Leighton dynamo model, the polar magnetic field of the
Sun around solar minimum should be a predictor for the strength of the
next cycle: since for such models the polar field is thought to reflect
the global dipolar poloidal field from which the toroidal field of the
next cycle is being generated by differential rotation, the strength (or
magnetic flux) of this toroidal field should by higher for a stronger
polar field. Since measurements of the polar fields are rather uncertain
and consistent data series are available only since a few decades, a
stringent test of the suggestion of \citet{Schatten:etal:1978} could not
be carried out so far. Various proxies for the polar field have also
been considered, but with inconclusive results
\citep{Layden:etal:1991}. Nevertheless, the method has been used to
predict a rather weak solar cycle 24
\citep{Svalgaard:etal:2005,Schatten:2005}.

Although they refer to dynamo theory (in fact, to a vague notion of the
Babcock-Leighton model), the proponents of the polar field precursor
have never actually used a mathematical dynamo model to support their
suggestion, neither in the original paper \citep{Schatten:etal:1978}
nor in any of the follow-up papers. In a very crude way, such an
attempt has been made only very recently by
\citet{Choudhuri:etal:2007}. These authors use a Babcock-Leighton-type
flux-transport dynamo model and arbitrarily rescale the poloidal field
at 4 cycle minima according to measured polar field values (from the
Mount Wilson and Wilcox solar observatories). It is not surprising (in
fact, almost trivial) that the toroidal fields of the respective
following cycles reflect the value of the scaling factor. In fact, any
linear or mildly nonlinear model would lead to the same result, so that
\citet{Choudhuri:etal:2007} effectively do not go beyond the original
suggestion of \citet{Schatten:etal:1978}. This is also demonstrated by
\citet{Brandenburg:Kaepylae:2007} who obtain practically the same result
with a heavily truncated toy model. Therefore, such a crude
approach to `data assimilation' does not provide more information than
simple correlation studies and, in particular, does not furnish
constraints for dynamo models.

The approach of \citet{Dikpati:etal:2006} and
\citet{Dikpati:Gilman:2006}, hereafter referred to as the DDG model, is
the first serious attempt to use a mean-field dynamo model to predict
solar cycle strength. These authors use an axisymmetric
(longitude-averaged) flux-transport dynamo model in a spherical shell
with a solar-like meridional flow (poleward at the surface) and a low
turbulent diffusivity in the convection zone. The differential rotation
is chosen according to the helioseismic measurements; it generates
toroidal magnetic flux near the bottom of the convection zone, the
amount of which which is taken as the predictor quantity. In such
models, the dynamo loop is usually closed by assuming an $\alpha$-effect
relating the toroidal field to a near-surface source term for the
poloidal field. In the DDG model, this kind of closure is replaced by a
source term that directly reflects the observed emergence of tilted
bipolar magnetic regions: the source with Gaussian latitude profile
drifts between 35 deg and 5 deg latitude during a sunspot cycle and its
strength is scaled with the historical record of observed sunspot areas
since 1876 (RGO data plus extensions since 1976). Through this data
assimilation procedure, the source term reflects the actual variations
of the flux emergence at the surface and incorporates them into the
evolution of the model. The DDG model provides amazingly high
correlation coefficients between the amount of low-latitude toroidal
magnetic flux in the deep convection zone calculated (`predicted') by
the model and the strength (maximum sunspot number) of the corresponding
cycle; values up to $r=0.99$ are obtained.

The success of the DDG model is surprising given the various assumptions
and parametrizations entering the model, for instance: (1) arbitrary
prescription of the (unknown) meridional flow pattern in the deep
convection zone, (2) a strong radial drop of the turbulent magnetic
diffusivity between the surface layers and the deeper parts of the
convection zone, (3) schematic prescription of the profile, width and
latitude drift of the poloidal field source. This has led
\citet{Bushby:Tobias:2007} to argue that the correlations obtained by
DDG are either fortuitous or the result of parameter tuning, claiming
that it is "impossible to predict the solar cycle using the output of
such models". They give two examples to support this claim: a) a
flux-transport model with stochastic fluctuations of the meridional
flow, and 2) an interface dynamo with a strong nonlinearity
(back-reaction on the differential rotation). While the arguments of
\citet{Bushby:Tobias:2007} certainly apply to their kind of "ab-initio"
dynamo models with a closed dynamo loop, it is not so clear how much
weight they carry concerning the data assimilation approach of DDG: in
fact, using the observed flux emergence takes account of at least part
of the random fluctuations and nonlinearities certainly inherent in
working of the solar dynamo, namely those associated with the connection
between the toroidal field deep in the convection zone and the surface
field.  Precisely these variations in the source strength eventually
determine the modulation of the cycle amplitudes in the DDG model, but
in a non-trivial way (as exemplified by the correctly reproduced drop of
activity from cycle 19 to 20, in spite of a strong source amplitude
provided by the flux emergence in cycle 19). Other fluctuations, such as
variations of the meridional flow, could also be incorporated into the
model once sufficiently detailed and extended measurements become
available.

Even if the claims of \citet{Bushby:Tobias:2007} would apply to
the DDG model, the question remains why this model provides such
high correlations. Could it really be parameter tuning? This can be
tested in a straightforward manner: take the DDG model and use a source
with random amplitudes for 12 cycles; then tune the model parameters
such that the predictor (toroidal field) reproduces the actual maxima of
the last 9 cycles (with a correlation coefficient exceeding 0.95,
say). I very much doubt that the DDG model has such a degree of
flexibility and I would dare to predict that this will turn out to be an
impossible task indeed!

So, after all, the DDG model cannot be brushed away off-handedly.  We
need to understand where its predictive skill comes from, since this
might tell us something important about the solar dynamo: for instance,
does the evolution of the surface flux during a cycle play a crucial
role in the dynamo process (and affect the strength of subsequent
cycles) or is it just a superficial epiphenomenon of the hidden dynamo?

\section{A simple flux transport model}

Given its parametrization of poorly know properties (such as internal
meridional flow and turbulent diffusivity), is it conceivable that the
details of the interior in fact do not matter for the correlations
obtained by DDG? If that would be the case, then a pure surface
transport model driven by the same source (emerging flux) as used in the
DDG model should already contain and reveal the relevant information.
In a recent paper \citep{Cameron:Schuessler:2007}, we have therefore
considered a very simple (almost trivial) axisymmetric surface flux
transport model for the radial magnetic field component as a function of
latitude and time. Flux is fed into the system by a source term
analoguous to the DDG source and we follow its subsequent evolution
under the influence of a poleward meridional flow and turbulent
diffusion. We have considered various quantities as predictors. In the
spirit of the Babcock-Leighton model, the amount of magnetic flux
diffusing over (or reconnecting at) the equator is the most
relevant quantity: only this part of the emerged flux represents the
global dipole field that acts as the poloidal field source for the
toroidal field of the next cycle. If we match as closely as possible the
procedures and parameter choices in the DDG model, we indeed find
that the cross-equator flux during cycle $n$ is correlated with the
maximum sunspot number of cycle $n+1$ with $r=0.9$. The result turns out
to be fairly robust with respect to parameter variations; values up to
$r=0.95$ can be reached by `tuning'. Incidentally, taking the polar field
strength as a predictor results in much lower correlation coefficients. 

\begin{figure}
 \centering
 \includegraphics[angle=-90,width=0.9\linewidth]{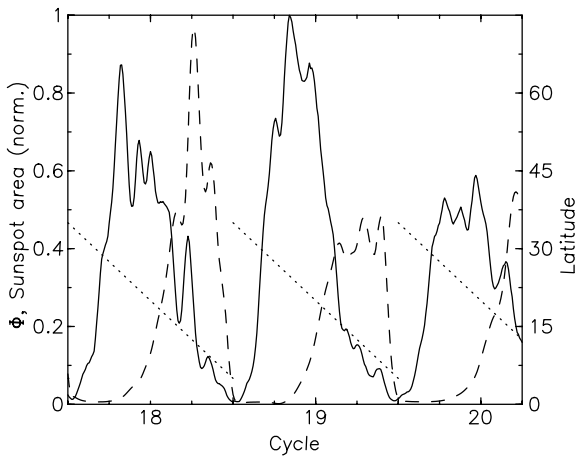}
 \caption{Observed sunspot area (solid curve) and rate of magnetic flux
  diffusing over the equator (dashed curve), both normalized, for solar
  cycles 18, 19, and 20. The dotted lines (scale to the right) indicate
  the latitude drift of the centroid of the Gaussian source term
  representing flux emergence (progression of the sunspot
  belt). The cross-equator flux peaks during the decling phase of the
  cycle, a few years before the minimum epochs, when the source has
  reached low latitudes \citep[from][]{Cameron:Schuessler:2007}.}
\label{fig:blowup}
\end{figure}

How do the high correlation doefficient with the cross-equator flux come
about? For instance, why is the comparatively weak cycle 20 correctly
`predicted' although a large amount of surface flux has emerged during
the preceding cycle 19?  Fig.~\ref{fig:blowup} explains how this
works. The cross-equator magnetic flux peaks a few years before sunspot
minimum during the time when the source flux is supplied at low
latitudes, so that 1) the distance to the equator is smaller and 2) the
meridional flow (antisymmetric to the equator) is less forceful. Both
effects lead to a a strongly increasing rate of magnetic flux diffusing
over the equator when the source approaches the equator.  As a
consequence, the amplitude of the predictor is determined by the amount
of flux emergence (sunspot area) in the declining phase of the
cycle. The relatively high flux emergence during the declining phase of
cycle 18 leads to a large value of the predictor for cycle
19. Significantly less flux erupted during this phase in cycle 19, so
that the predictor for cycle 20 is lower, although the total flux
emergence during cycle 19 is much higher than that of cycle 18.  This
result already indicates that the predictor could be rather sensitive to
the definition of the source latitudes in the model.

Having understood that the level of activity a few years before minimum
mainly determines the amplitude of the predictor in our simple
surface-transport model, we may ask ourselves whether we need the model
at all in order to make a prediction.  In fact, taking the level of
recorded sunspot number three years before minimum and correlating it
with the strength of the next maximum for all cycles since 1750 leads to
a value of $r=0.89$ for the correlation coefficient.

On the other hand, can we possibly improve the flux-transport model?  We
have detailed information about the areas and latitudes of individual
sunspot groups for the whole period since 1874 \citep[RGO, SOON and
Russian data, see][]{Balmaceda:etal:2005}, so that we can replace the
schematic source term of DDG by a procedure that separately takes into
account each sunspot group in the data. The surprising result is a
dramatic drop of the correlation between the cross-equator flux and the
strength of the following cycle: with $r=0.33$, the predictive skill
is almost gone. In fact, the predictor now correlates better with the
strength of the ongoing cycle than with the next cycle.

\section{The origin of the predictive skill}

The results sketched in the previous section leave us with puzzling
questions. Why is there predictive skill in the flux-transport model
with the schematic source and why does it completely vanish for more
realistic input data? Why does the 3-year precursor based upon sunspot
numbers work reasonably well? Has any of this anything to do with the
Babcock-Leighton dynamo scheme?

With the benefit of hindsight, the answer to these questions seems
amazingly simple, almost trivial. Let us first remind ourselves that
there is a third possibility for the origin of predictive skill in
linear models like the DDG approach or ours: besides 1) intrinsic
validity of the model and 2) sheer luck or parameter tuning, there could
be 3) correlations in the input data themselves. We shall see below that
such correlations indeed exist and that they probably are responsible
for the correlations obtained with most precursor methods and also with
our simple flux-transport model.
 
\begin{figure}[ht!]
 \centering
 \includegraphics[width=0.9\linewidth]{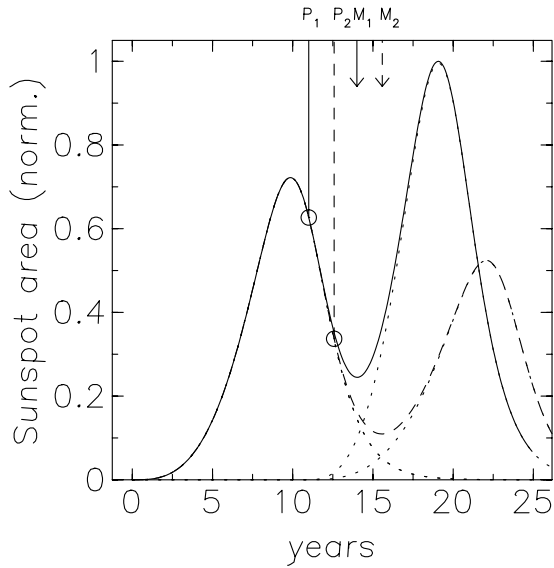}
 \caption{Schematic illustration of the amplitude-dependent shift of the
  minimum between overlapping, asymmetric sunspot cycles and its
  influence on a precursor quantity (sunspot activity 3 years before
  minimum). A stronger follower cycle (solid curve) with a shorter rise
  time leads to an earlier minimum (M1) and a higher predictor (P1) than
  a weaker subsequent cycle (dashed curve, minimum M2, predictor P2)
  with a longer rise time. Both alternatives for the follower cycle
  start at $t=11\,$yr \citep[from][]{Cameron:Schuessler:2007}.
  }
\label{fig:idea}
\end{figure}

It turns out that a combination of two well-known properties of the solar
cycle explains (or, at least, contributes a significant part to) the
predictive skill of precursor-type models:
\begin{enumerate}
\item {\em overlapping of cycles:} active regions belonging to the
  new cycle start to appear in mid latitudes while there is ongoing flux
  emergence  near the equator connected with the old cycle.
\item {\em Waldmeier's rule:} stronger cycles tend to rise faster towards
  sunspot maximum \citep{Waldmeier:1935}.
\end{enumerate}
The important point is that both properties make the level and the
timing of the formal solar minimum (epoch of minimum sunspot number)
depend on the strength of the following cycle. This is exactly the
correlation in the sunspot number (or area) data that eventually leads
to the predictive skill of precursors.  Fig.~\ref{fig:idea} illustrates
schematically how this comes about. Given are time profiles of
overlapping sunspot cycles according to an empirical functional form
that reproduces both the rise and decay parts of a cycle, including
Waldmeier's rule \citep{Hathaway:etal:1994,Li:1999}. The figure shows
the effect of the strength of the following cycle (dotted curves) on the
summed activity levels around minimum activity between the cycles.  The
faster rise of a stronger follower cycle leads to an earlier and higher
sunspot minimum in the summed activity curve (solid line) than in the
case of a weaker follower cycle (dashed line). In the case shown, the
time shift of the minimum epoch is about one year. Since a sunspot cycle
is defined as the time between adjacent minima, the activity in the
declining phase of the first cycle, (i.e., in a fixed time interval
relative to the respective solar minimum epoch) is considerably larger
when the follower cycle is stronger than when it it weaker. When a
precursor is taken relative to sunspot minimum (e.g., our choice of the
sunspot number 3 years before minimum), it is obvious that its level
indeed will reflect the strength of the following cycle -- without
requiring any kind of direct physical connection between the precursor
and the following cycle amplitude.

It is clear that the correlation in the input data (sunspot area record)
explained above also underlies the predictive skill of our
flux-transport model with a schematic source. In this case, we have
assumed \citep[following][]{Dikpati:etal:2006} a fixed latitude
progression of the source centroid from 35 deg to 5 deg between two
sunspot minima. Consequently, in the case of a strong follower cycle,
higher activity levels in the descent phase (a few years before
minimum) due to the correspondingly earlier minimum epoch are mapped to
lower emergence latitudes and, therefore, lead to a higher amount of
magnetic flux diffusing over the equator. If we directly take the
emerging active regions with their actual emergence latitudes from the
available record, then the near-equator flux emergence comes only from
the preceding cycle, and thus correlates with its strength. As a
consequence, our precursor does not show predictive skill any more and,
in fact, only reflects the strength of the `old' cycle. We are not in a
position to claim that the same explanation also holds for the
correlations found with the DDG model, but this can be tested by
replacing their schematic source term by the existing sunspot group data
with actual emergence latitudes.

Apart from explaining the predictive skill of a many precursor
quantities measured during the descent phase or around solar minimum,
the overlapping of cycles and Waldmeier's rule also naturally accounts
for a number of well-known correlations in the sunspot record, for
instance: 1) strong cycles tend to be preceded by short cycles
\citep[e.g.,][]{Solanki:etal:2002b}, 2) minimum levels preceding strong
cycles tend to be higher \citep{Hathaway:etal:2002}, and 3) more asymmetric
cycles tend to be followed by weaker cycles
\citep[e.g.,][]{Lantos:2006}. 

\section{Conclusions}

I think that the correlations introduced into the sunspot number and
sunspot area records by the combination of cycles overlap and
Waldmeier's rule go a long way towards explaining the predictive skill
of many precursor approaches as well as the correlations provided by the
flux-transport models with a schematic source. Consequently, there is
more to these models than just numerology or parameter-tuning. However,
the key point is not so much to predict but to {\em understand} the
solar cycle.  So what have we learned in this respect? Not very much, I
am afraid: the correlation introduced by the Waldmeier effect of
overlapping cycles does not require any kind of physical relation
between the surface fields of the previous cycle(s) and the strength of
the following cycle; in particular, it cannot be taken as evidence in
favour of a Babcock-Leighton type dynamo model. In fact, it can be shown
that precursor methods successfully predict cycle sequences with
randomly varying strength \citep{Cameron:Schuessler:2007}. On the other
hand, these results do not exclude a physical connection between
precursor and following cycle strength. For instance, the precursors
could also be affected by flux emergence in high latitudes, e.g., in the
form of ephemeral regions preceding the appearance of the first sunspots
of the new cycle \citep[e.g.,][]{Harvey:1993,Harvey:1994a}, so that the
new cycle would already directly affect the surface flux during the
descending phase of the old cycle. These all remain valid possibilities,
it is only that the predictive skill of precursor methods per se does
not help us to decide which of these is in fact realised by the Sun.

In all such considerations we should not forget that all the
relationships that may be used for prediction are `noisy' and thus valid
only in a statistical sense. The existence of grand minima like the
Maunder minimum reminds us that the Sun has much more variability in
store than simple statistical analysis of sunspot data would be able to
predict. And even if a prediction method has a good correlation record
for the past, it may completely fail for the next cycle. The split
opinion of the NOAA/NASA Solar Cycle 24 Prediction panel\footnote{see
http://www.sec.noaa.gov/SolarCycle/SC24} about whether the coming cycle
would be high or low provides a good illustration about the `state of
the art' -- and may actually reflect intrinsic limitations as
illustrated by the examples given by \citet{Bushby:Tobias:2007}.

So, where do we stand now concerning the question in the title? We have
seen that, owing to the cycle overlap and the Waldmeier effect,
predictor methods can obtain relevant information about the new cycle at
the epoch around solar minimum. However, the underlying statistical
relationships contain a significant amount of scatter, so that actual
predictions are uncertain, as their mixed performance in the past
clearly shows. The skill of such predictor schemes does not seem to
provide constraints or relevant information about the working of the
solar dynamo, apart from the trivial fact that a valid dynamo model
ultimately will have to reproduce and explain the underlying
correlations. Ever since reading the paper of
\citet{Legrand:Simon:1981}, I had hoped that there would be more to
learn.

In order to end on a more positive note, let me say that, of course, the
last word on these matters is not spoken. It may turn out, after all,
that the data assimilation model of \citet{Dikpati:etal:2006} and
\citet{Dikpati:Gilman:2006} will pass with flying flags the crucial test
of using the actual flux emergence events in its source - and that these
results will be independently confirmed by others and without excessive
parameter tuning. Then sceptics like \citet{Bushby:Tobias:2007} or
\citet{Cameron:Schuessler:2007} would have a hard time to search for
sources of the predictive skill other than the operation of a
Babcock-Leighton-type dynamo. It would also be worthwhile to look for
signatures of the new cycle during the post-maximum phase (or even
before) of the ongoing cycle, e.g., by monitoring flux emergence and the
evolution of large-scale magnetic patterns in mid/high latitudes and
compare with surface flux-transport simulations.  Observational data
from SOLIS, Hinode, SDO and eventually Solar Orbiter will be
particularly suitable for this purpose. A positive detection of such
signatures could possibly extend the lead time for solar cycle
prediction using precursors.

%\acknowledgements

%\bibliographystyle{apj}
%\bibliography{journals,papref}

\bibliography{pap.bbl}

%\begin{thebibliography}{}
%\end{thebibliography}

\end{document}